\documentclass[fleqn,usenatbib]{mnras}

\usepackage{newtxtext,newtxmath}
\usepackage{multirow}

\usepackage[T1]{fontenc}

\DeclareRobustCommand{\VAN}[3]{#2}
\let\VANthebibliography\thebibliography
\def\thebibliography{\DeclareRobustCommand{\VAN}[3]{##3}\VANthebibliography}

\usepackage{graphicx}	
\usepackage{amsmath}	

\graphicspath{ {./Thesis/Thesis_Figures/} {./MG_MSW/} {./SG_MSW/} }

\title[Shock Cooling Model: III BSG's]{Shock cooling emission from explosions of massive stars: III.  Blue Super Giants}

\author[Morag et al.]{
Jonathan Morag\thanks{E-mail: jonathan.morag@weizmann.ac.il},$^{1}$
Nir Sapir,$^{1,2}$
Eli Waxman$^{1}$
\\
$^{1}$Weizmann Institute of Science, Rehovot, Israel\\
$^{2}$Soreq Nuclear Center, Yavneh, Israel\\
}

\date{Accepted XXX. Received YYY; in original form ZZZ}

\pubyear{2022}

\begin{document}
\label{firstpage}
\pagerange{\pageref{firstpage}--\pageref{lastpage}}
\maketitle

\begin{abstract}

Light emission in the first hours and days following core-collapse supernovae is dominated by the escape of photons from the expanding shock-heated envelope. In preceding papers, we provided a simple analytic description of the time-dependent luminosity, $L$, and color temperature, $T_{\rm col}$, as well as of the small ($\simeq10\%$) deviations of the spectrum from blackbody at low frequencies, $h\nu< 3T_{\rm col}$, and of  `line dampening' at $h\nu> 3T_{\rm col}$, for explosions of red supergiants (RSGs) with convective polytropic envelopes (without significant circum-stellar medium). Here, we extend our work to provide similar analytic formulae for explosions of blue supergiants with radiative polytropic envelopes. The analytic formulae are calibrated against a large set of spherically symmetric multi-group (frequency-dependent) calculations for a wide range of progenitor parameters (mass, radius, core/envelope mass ratios) and explosion energies. In these calculations we use the opacity tables we constructed (and made publicly available), that include the contributions of bound-bound and bound-free transitions. They reproduce the numeric $L$ and $T_{\rm col}$ to within 10\% and 5\% accuracy, and the spectral energy distribution to within $\sim20-40\%$. The accuracy is similar to that achieved for RSG explosions.

\end{abstract}

\begin{keywords}
radiation: dynamics – shock waves – supernovae: general
\end{keywords}



\section{Introduction}

In core collapse supernovae (SNe) explosions, a radiation mediated shock (RMS) traverses outwards through the stellar progenitor, heating and expelling material as it passes. If no significant circumstellar material (CSM) is present around the star, arrival of the shock at the surface produces a hard UV/X-ray $\sim10^{45} \, \rm erg \, s^{-1}$ `shock-breakout' emission, lasting from tens of minutes to an hour. The breakout pulse is then followed in the coming hours and days by thermal UV/optical `shock-cooling' emission, caused by diffusion of photons out of the shock-heated stellar ejecta. Typical luminosities and temperatures during shock-cooling are of the order $10^{42}-10^{44} \rm \, erg \, s^{-1}$, and $1-10$ eV. As the photons diffuse out, deeper parts of the ejecta are gradually exposed over time \citep[see][for reviews]{waxman_shock_2016,levinson_physics_2020}.

\defcitealias{rabinak_early_2011}{RW11}
\defcitealias{sapir_uv/optical_2017}{SW17}

In order to constrain the properties of the progenitor star, it is helpful to have high cadence multi-band observations in the first hours of shock-cooling \citep[see, e.g. ][and references therein]{morag_shock_2023}. Among these measurements, ultraviolet observations are especially important, as they are closer to the thermal emission peak, and can be used to determine the emission temperature and the UV extinction self-consistently \citep[][the first two hereafter \citetalias{rabinak_early_2011} and \citetalias{sapir_uv/optical_2017}]{rabinak_early_2011,sapir_uv/optical_2017,sagiv_science_2014,rubin_exploring_2017}.
Combined with an accurate theoretical model, these measurements can be used to reproduce the progenitor and explosion parameters, including radius, surface composition, explosion energy per unit mass, and the extinction. Analytic models are especially important for solving the "inverse problem" of inferring system parameters and uncertainties from the observed spectral energy distribution (SED).

Catching supernovae within the first hour presents a practical challenge, and few such observations have been achieved \citep[see discussions in][]{morag_shock_2023,irani_sn_2022}. Existing and upcoming observatories, such as the Zwicky Transient Factory (ZTF) \citep{gal-yam_real-time_2011}, the upcoming Vera Rubin Observatory \citep{ivezic_lsst_2019}, and the expected launch of the wide-field UV space telescope ULTRASAT \citep{sagiv_science_2014,shvartzvald_ultrasat_2023} will greatly increase the quantity and quality of early measurements, enabling a systematic study.

\defcitealias{sapir_non-relativistic_2011}{SKW I}
\defcitealias{katz_non-relativistic_2012}{KSW II}
\defcitealias{sapir_non-relativistic_2013}{SKW III}
\defcitealias{nakar_early_2010}{NS10}
\defcitealias{shussman_type_2016}{SWN16}
\defcitealias{piro_shock_2021}{PHY21}
\defcitealias{morag_shock_2023}{Paper I}
\defcitealias{morag_shock_2024}{Paper II}

Emission during shock-cooling is amenable to accurate modeling in the case of `envelope breakout' (i.e. in the absence of CSM with significant optical depth), largely because the system is near local thermal equilibrium (LTE) at this time, the hydrogen is highly ionized, and the opacity is dominated over much of the frequency range by electron scattering. As explained in our earlier papers, our results are accurate up to the Hydrogen recombination time, roughly when $T_{\rm col}$ drops to 0.7~eV. In preceding papers, \citet[][hereafter \citetalias{morag_shock_2023} and \citetalias{morag_shock_2024} respectively]{morag_shock_2023,morag_shock_2024}, we provided analytic formulae describing the shock-cooling emission following core-collapse in red supergiant stars (RSG). A recent study of a large set of type II SN observations \citep{irani_sn_2022} finds that our model accounts well for the early multi-band data of 50\% of observed SNe, corresponding to 70\% of the intrinsic SN distribution after accounting for luminosity bias (the others are likely affected by thick CSM). This agreement enables the inference of progenitor radius, explosion velocity, and relative extinction by using the formulae provided in \citetalias{morag_shock_2023} and \citetalias{morag_shock_2024}.

In this paper, we extend our work to provide similar analytic formulae for explosions of blue supergiants (BSG) with radiative polytropic envelopes. The analytic formulae are calibrated against a large set of 1D multi-group (frequency-dependent) calculations for a wide range of progenitor parameters: radii in the range $R=10^{12}-3\times10^{13}$~cm, explosion energies in the range $E=10^{50}-10^{52}$~erg, core/envelope mass ratios in the range $M_{\rm c}/M_{\rm e}=0.1-3$. We have demonstrated in earlier work \citetalias{sapir_uv/optical_2017,morag_shock_2023}, that the shock-cooling emission characteristics are insensitive to metallicity, to deviations from ideal `polytropic' structure, and to core radius variations. We, therefore, do not explore the dependence on these parameters in this paper. Our numeric calculations use the opacity tables we constructed (and made publicly available), that include the contributions of bound-bound and bound-free transitions.

The paper is structured as follows. In \S~\ref{sec: Earlier Results} we summarize the analytic results of earlier work that we use in this paper. We concisely describe our numerical code and opacity tables in \S~\ref{sec: numerical code}. In \S~\ref{sec:analytic model}, we provide our calibrated analytic formulae, and in  \S~\ref{sec: numeric results} we assess their accuracy by comparisons to numeric results. In \S~\ref{sec: expansion opac} and \S~\ref{sec: NLTE}, we show that our results are not sensitive to the effects of "expansion opacity" and to deviations from LTE excitation and ionization. In \S~\ref{sec: STELLA compare} we compare our results to the results of similar STELLA simulations. The results are summarized and discussed in \S~\ref{sec: conclusion}. A complete description of our analytic formulae for both BSG and RSG progenitors is given in the appendix.

\section{Earlier Analytic Results}
\label{sec: Earlier Results}

We summarize below the analytic results of earlier work that we use in this paper. In the case of a polytropic density profile, the density near the stellar edge is given by
\begin{equation}
\label{eq:rho_in}
    \rho_0 = f_\rho \bar{\rho} \delta^n,
\end{equation}
where $\delta \equiv (R-r)/R$, $r$ is the radial coordinate, the total average progenitor density is $\bar{\rho}\equiv M/(4\pi R^3/3)$, $f_\rho$ is a constant of order unity, and $n=3/2,3$ for convective and radiative envelopes, respectively. To avoid confusion, all values dependent on the choice of $n=(3/2,3)$ are provided in this paper for BSG's $(n=3)$ only, except for the appendix that summarizes the formulae for both cases.

Following core-collapse, the outward-propagating shock accelerates down the steep density gradient, with shock velocity increasing in a self-similar manner according to \citep{gandelman_shock_1956,sakurai_problem_1960}
\begin{equation}
\label{eq:vs}
    \rm v_{\rm sh} = v_{\rm s\ast} \delta^{-\beta_1 n},
\end{equation}
where $\beta_1=0.19$, and $\rm v_{\rm s\ast}$ is approximately given by  \citep{matzner_expulsion_1999}
\begin{equation}
\label{eq:vstar}
    {\rm v_{\rm s\ast}}\approx 1.05 f_\rho^{-\beta_1}{\rm v_\ast,\quad v_\ast}\equiv\sqrt{E/M}.
\end{equation}

Photons from the RMS will escape the shock and the star, i.e. will "breakout" when the scattering optical depth ahead of the shock approaches the optical depth across the width of the shock, $\tau = c/{\rm v}_{\rm sh}$ \citep{ohyama_explosion_1963}. This occurs at
\begin{equation} \label{eq:delta_bo_def}
    \delta_{\rm bo} = (n+1)\frac{c }{ \kappa \rho_{\rm bo} {\rm v_{\rm bo}} R},
\end{equation}
where $\kappa$ is the opacity and the shock velocity and pre-shock density at breakout, $v_{\rm bo}$ and $\rho_{\rm bo}$, are defined by
\begin{equation}\label{eq:rho_v_0_def}
  \rho_0=\rho_{\rm bo} ({\rm v_{\rm bo}}\tau/c)^{n/(1+n)},\quad
  {\rm v_{\rm sh}} = {\rm v_{\rm bo} (v_{\rm bo}}\tau/c)^{-\beta_1 n/(1+n)},
\end{equation}
where $\tau$ is the scattering optical depth from $r$ to the stellar edge, $r=R$. For BGS's,
\citep{waxman_shock_2016}
\begin{align} \label{eq:rho_v_0_approx}
  \rho_{\rm bo} & = 5.6 \times 10^{-9} M_{0}^{0.13} {\rm v_{\ast,8.5}}^{-0.87} R_{12}^{-1.26} \kappa_{0.34}^{-0.87} f_{\rho}^{0.29}\, \rm g \, cm^{-3}, \nonumber\\
  {\rm v_{\rm bo}/v_{\ast}} & = 8.6 M_{0}^{0.16} {\rm v_{\ast, 8.5}}^{0.16} R_{12}^{-0.32} \kappa_{0.34}^{0.16} f_{\rho}^{-0.05},
\end{align}
and
\begin{equation}
\label{eq:dbo}
    \delta_{\rm bo}=0.088 \, R_{13}^{0.58} (f_\rho M_0 \, {\rm v_{\rm s*,8.5}}\, \kappa_{0.34})^{-0.29}.
\end{equation}
Here, $R= 10^{12}R_{12}$~cm, $\kappa=0.34 \kappa_{0.34} \rm \, cm^2 g^{-1}$, ${\rm v_\ast=v_{\ast,8.5}} 10^{8.5} \, \rm cm s^{-1}$, ${\rm v_{\rm s*}=v_{\rm s*,8.5}} 10^{8.5}$, and $M=1 M_{0} M_\odot$. 

The duration of the breakout pulse is related to the shock crossing time of the breakout layer as
\begin{equation}
\label{eq:tbo}
   \frac{\delta_{\rm bo}R}{\rm v_{\rm bo}} =\frac{(n+1)c }{ \kappa \rho_{\rm bo} v_{\rm bo}^2}= (n+1)t_{\rm bo}=350\, \rho_{\rm bo,-9}^{-1}\kappa_{0.34}^{-1}{\rm v_{\rm bo,9}}^{-2}{\, \rm s},
\end{equation}
where $\rho_{\rm bo}=10^{-9}\rho_{\rm bo,-9}{\rm \, g \, cm^{-3}}$ and $\rm v_{\rm bo}= 10^{9} v_{\rm bo,9} \,cm\,s^{-1}$. The observed pulse duration may be longer due to light travel time effects, which "smear" the pulse over $t\sim R/c$.

For later use we recast equation (9) of \citetalias{rabinak_early_2011} in terms of $r$ and $t$,
\begin{equation}
    \rho(r,t) = 1.26 \times 10^{-11}\, f_{\rm \rho} M_0 \, v_{\rm s*,8.5}^{7.18} r_{14}^{-10.18} t_{\rm d}^{7.1 8} \, \rm g \, cm^{-3},
\label{eq:rho_rt_vsstar}
\end{equation}
where$r=r_{14} 10^{14}$ cm. Alternatively, using equation \ref{eq:rho_v_0_def},
\begin{equation}
    \rho(r,t) = 3.64 \times 10^{-12} \, R_{13}^2 {\rm v_{\rm bo,9}^{6.18}} \kappa_{0.34}^{-1} r_{14}^{-10.18} t_{\rm d}^{7.18} \, \rm g \, cm^{-3}.
    \label{eq:rho_rt_bo}
\end{equation}
Using the self-similar diffusion profile of \citet{chevalier_early_1992} we have
\begin{equation}
    T(r,t) = 4.66 \, R_{13}^{1/4} (f_{\rho}M_0)^{0.29} {\rm v_{\rm s*,8.5}^{2.25}} \kappa_{0.34}^{0.04} t_{\rm d}^{1.71} r_{14}^{-2.79} \, \rm eV,
    \label{eq:T_rt_vsstar}
\end{equation}
\begin{equation}
    T(r,t) = 2.91 \, R_{13}^{0.75}  {\rm v_{\rm bo,9}^{1.88}} \rho_{\rm bo,9}^{-0.08} \kappa_{0.34}^{-1/3} t_{\rm d}^{1.71} r_{14}^{-2.79} \, \rm eV.
    \label{eq:T_rt_bo}
\end{equation}

Assuming a blackbody spectral distribution, the emitted luminosity is then given by,
\begin{equation}
    L_{\rm BB}=L\times\pi B_{\nu}(T_{\rm col})/\sigma T_{\rm col}^{4}.
    \label{eq:L_nu_BB_formula}
\end{equation}
In \citetalias{morag_shock_2023} we provided a slight modification to the blackbody formula, taking into account the modification due to line suppression in the ultraviolet,
\begin{equation}
        L_\nu = L \times \min \left[ \, \frac{\pi B_\nu(T_{\rm col})} {\sigma T_{\rm col}^4} \, , \, \frac{\pi B_\nu(0.74 T_{\rm col})} { \sigma (0.74 T_{\rm col})^4} \right].
 \label{eq:L_nu_BB_mod_formula}
\end{equation}

\section{Numerical Code}
\label{sec: numerical code}
Our gray and multigroup numerical codes are described in detail in \citetalias{morag_shock_2023} and \citetalias{morag_shock_2024}, along with multiple tests of the codes comparing the numeric results to analytic ones, e.g. for the steady planar RMS and shock breakout problems \citep[additional test problems in][]{sapir_numeric_2014}. We provide below a brief description of these codes and of our numercial analysis method, which are identical to those used (and described in detail) in \citetalias{morag_shock_2024}.

We solve the one-dimensional spherical Lagrangian hydrodynamics equations for ideal fluid flow coupled with radiative transfer under the diffusion approximation. Our gray simulations assume radiation pressure dominated flow and solve diffusion using constant electron scattering opacity, $\kappa_{\rm es}=0.34 \, \rm g \, cm^{-2}$. The color temperature of the emitted thermal flux from the gray simulations is determined in post-processing using Rosseland mean opacity as described in \S~\ref{sec:Tc}. In the multigroup simulations, we calculate both the plasma energy density $e$ and the frequency binned photon energy density $u_\nu$. We include frequency-dependent radiative emission/absorption and diffusion. For emission/absorption, we use a frequency-binned average of the opacity extracted from the high-frequency-resolution table. For the diffusion, we use a binned Rosseland mean. For all simulations, we place a static reflective boundary at the inner surface and a free boundary with an Eddington parameter at the outer edge.

The numerical analysis is carried out using a succession of simulations, with each simulation starting later in time using a snapshot of previous simpler simulations. We begin a hydrodynamics-only simulation of a "thermal bomb" placed near the center of a simplified progenitor star with a uniform high-density core and radiative polytropic envelope structure. When the shock reaches a scattering optical depth $\tau=(10-24)c/{\rm v_{\rm sh}}$, we begin a gray diffusion simulation. Later, at time $t=(1-2)R/c$ following breakout, we begin a multigroup simulation based on the instantaneous gray diffusion simulation profiles. All simulations are carried out until past the validity time (of H recombination) for later comparison.

Our Lagrangian spatial resolution grid is as described in \citetalias{morag_shock_2024}, with 4000-8000 cells in the hydrodynamic only simulations, 1600-3200 cells for the gray diffusion simulations, 50-100 spatial cells\footnote{the latter do not describe the breakout pulse and as a result have lower resolution requirements} and up to 256 photon groups for multigroup simulations. As before, we check for convergence, though based on convergence results of earlier calculations, we do this only for a sub-sample of the simulations. Changing both the optical depth of the outermost cell from $\tau=10^{-2}$ to $10^{-3}$ and the cell count from 50 to 100 leads to small, 1-few percent deviations, with 10\% deviations for some frequencies in extreme cases with no effect on the calibration of our analytic formulas.

Following breakout, the instantaneous velocity of the ejecta is monotonically increasing outwards except for the breakout shell, where the velocity profile is flat or even mildly decreasing. The decreasing profile near the edge leads to the formation of a high density shell (which may be unstable), which does not affect the escaping radiation significantly but can lead to numerical issues. BSG's exhibit a more pronounced density inversion, as the velocity decrease can be up to $\sim10\%$ of the maximum velocity (compared with $0.1\%$ for RSG's). To mitigate numerical issues, as in \citetalias{morag_shock_2023} and \citetalias{morag_shock_2024}, we add an artificial viscosity term $q$ proportional to the velocity difference across cells. Its strength is dependent on spatial resolution, but it always satisfies $q\ll u$, where $u$ is the energy density. In a few cases of our analysis, we combine gray simulations at low spatial resolution (50 cells - largest $q$) with our higher resolution gray sims (1600-3200 cells), specifically in Fig. \ref{fig:L_RMser}. Our quoted RMS agreement between the analytic formula and simulation is unaffected by this choice of resolution.

\subsection{Opacity table}
Our frequency-dependent opacity table is calculated assuming LTE plasma ionization and excitation and includes free-free, bound-free, and bound-bound components. The latter is based on the Kurucz line list \citep{kurucz_atomic_1995} that is experimentally verified near recombination temperatures. In \citetalias{morag_shock_2023} and \citetalias{morag_shock_2024}, we showed that for pure Hydrogen and for solar mix compositions, our table provides similar results (Rosseland mean, frequency-dependent opacity) to those of TOPS, with the exception of bound-bound lines, which can be important for our problem. We also tested convergence of the results with respect to the resolution of the underlying density, temperature, and frequency grid, finding convergence using an underlying grid of $\Delta\nu/\nu\sim10^{-5}$. Our opacity code is publicly available for use on github \citep{morag_frequency_2023}.

When extracting the color temperature in the gray simulations, we use TOPS in our analysis for temperatures above $4$ eV, and our own table below 4 eV, as was done in \citetalias{morag_shock_2023}. In the multigroup simulations, we use our own opacity table and not TOPS. As discussed in  \citetalias{morag_shock_2023}, at later times it is more appropriate to use our own table, while at earliest times ($T>4$ eV) the observable UV/optical lightcurve is minimally affected by the presence of atomic lines. We previously showed weak sensitivity to metallicity so use here only solar mix values (comprised of 10 important elements up to Fe).

\subsection[Determining Tcol]{Determining $T_{\rm col}$ and $T_{\rm col,\nu}$ in post-processing}
\label{sec:Tc}
We repeat the procedures in \citetalias{morag_shock_2023} and \citetalias{morag_shock_2024} for extracting the color temperature by "post-processing" of our numerical results. For gray simulations, $T_{\rm col}(t)$ is obtained from the hydrodynamic profiles as the plasma temperature at the "thermal depth", $r_{\rm Th}(t)$, from which photons diffuse out of the envelope without further absorption. Following \citetalias{rabinak_early_2011} and \citetalias{sapir_uv/optical_2017}, we approximate $r_{\rm Th}(t)$ by 
\begin{equation}
\label{eq:Itais_Prescription}
\int_{r_{\rm Th}(t)}\rho\sqrt{3\kappa_{\rm R}\left(\kappa_{\rm R}-\kappa_{\rm es}\right)}dr'=1.
\end{equation}
Here, $\kappa_{\rm es}(\rho,T)$ is the electron scattering opacity, accounting for the ionization fraction, $\kappa_{\rm R}(\rho,T)$ is Rosseland mean opacity, and $(\kappa_{\rm R}-\kappa_{\rm es})$ represents the effective absorption opacity, $\kappa_{\rm abs}$.

For later use, we also define the frequency dependent color temperature $T_{\rm col,\nu}$ at the thermal depth $r_{\rm col,\nu}$ by
\begin{equation}
    \tau_{\star,\nu}(r=r_{{\rm col},\nu})\equiv\int_{r_{{\rm col},\nu}}^{\infty}\rho\sqrt{3\kappa_{\rm abs,\nu}\left(\kappa_{\rm abs,\nu}+\kappa_{\rm es}\right)}dr'=1,
    \label{eq: Thermal Depth Integral k_abs_nu}
\end{equation}
where the abs, es and $\nu$ subscripts indicate absorption, (electron) scattering and frequency dependence, respectively.

\section{Calibrated Analytic Model}
\label{sec:analytic model}
\subsection{Gray Formulae}
\label{sec:gray analytic model}
As was done in \citetalias{morag_shock_2023} for the gray formula in red supergiants, we combine the exact planar phase solutions of \citetalias{katz_non-relativistic_2012} with the approximate equations for the spherical phase of \citetalias{rabinak_early_2011}/\citetalias{sapir_uv/optical_2017} for BSG's. The result is very similar to equation (33) in \citetalias{morag_shock_2023}, with the addition of a weak dependency on progenitor radius,
\begin{equation}
    L(t)=L_{\rm planar} + L_{\rm RW} \times R_{13}^{0.1} A \, \exp\left[ -\left( at/t_{\rm tr} \right)^\alpha \right],
\end{equation}
\begin{equation}
    T(t) = \min(T_{\rm planar} , 1.07 \times T_{\rm ph,RW}).
\end{equation}
The values $\{A,a,\alpha\}=\{0.79,4.57,0.73\}$ are the same as given in \citetalias{sapir_uv/optical_2017}.
Analyzing our numeric results we find that for BSG's $v_* \sim 0.7 v_{\rm s,*}$ to $\pm15\%$ (recall that for RSG's the relation is $v_* \approx v_{\rm s,*}$).
 
We again simplify our analytic model, using an analytic approximation for the planar phase, 
\begin{equation}
\label{eq:L_gray_br}
    L/L_{\rm br}=\tilde{t}^{-4/3}+ A R_{13}^{0.1}\exp\left[-\left(at/t_{\rm tr}\right)^{\alpha}\right]
    \tilde{t}^{-0.34},
\end{equation}
and
\begin{equation}
\label{eq:Tcol_gray_br}
    T_{\rm col}/T_{\rm col,br}=\min\left[0.97\,\tilde{t}^{-1/3},\tilde{t}^{-0.48}\right],
\end{equation}
Here $\tilde{t}$ is the time normalized to the "break time," the time of transition from planar to spherical expansion,
\begin{equation}
    \label{eq:t-tild}
    \tilde{t}=t/t_{\rm br},
\end{equation}
with
\begin{equation}
\label{eq:t_br_of_vs}
    t_{\rm br}= 0.69 \, R_{13}^{1.32} v_{\rm s*,8.5}^{-1.16}
(f_{\rho}M_0\kappa_{0.34})^{-0.16}\,\text{hrs},
\end{equation}
\begin{equation}
\label{eq:L_br_of_vs}
L_{\rm br}=7.16\times10^{42} \, R_{13}^{0.55} v_{\rm s*,8.5}^{2.22}
(f_{\rho}M_0)^{0.22} \kappa_{0.34}^{-0.77}\,{\rm erg \, s^{-1}},
\end{equation}
\begin{equation}
\label{eq:T_br_of_vs}
T_{\rm col,br}= 9.49 \, R_{13}^{-0.38} v_{\rm s*,8.5}^{0.62}
(f_{\rho}M_0)^{0.07} \kappa_{0.34}^{-0.18}\,{\rm eV}.
\end{equation}

$L_{\rm br}$, $T_{\rm col,br}$ and $t_{\rm tr}$ can be directly deduced from observations, and their determination constrains the model parameters. For example, $R$ is given by
\begin{equation}
    R_{13}= 2.22 \, L_{\rm br,42.5}^{0.56} t^{-0.12}_{\rm br,3} T_{\rm br,5}^{-2.24},
    \label{eq:R_br}
\end{equation}
where $t_{\rm br}=3 t_{\rm br,3} \, \rm hours$, $L_{\rm br}=10^{42.5} L_{\rm br,42.5} \, \rm erg \, s^{-1}$, $T_{\rm br}=5 T_{\rm br,5} \, \rm eV$. 
For the BSG parameters' range we consider, $t_{\rm br}=30$ sec - 1 day, $L_{\rm br}=8\times10^{40}-3\times10^{43} \, \rm erg \, s^{-1}$, $T_{\rm br}=$ 1.5 - 32 eV (and $f_\rho M=0.2-40 \, M_{\odot}$, $v_{\rm s*,8.5}=$0.3-2.3).

Our analysis is valid, and the above formulae provide an accurate description of the emitted radiation, at
\begin{equation}
\label{eq:t_lc1}
   \max[2 R / c,t_{\rm bo}] < t < \min[t_{\rm 0.8 \, eV}, t_{\rm tr}/a],
\end{equation}
where for BSGs
\begin{equation}
    \label{eq: t_08eV}
    \begin{split}
        \begin{aligned}
        t_{\rm 0.8\,  eV} =& \,4.46 R_{13}^{0.52} v_{\rm s*,8.5}^{0.14} (f_{\rho}M_0)^{-0.02}\kappa_{0.34}^{-0.55} \, \rm days, \\
                    = & \,5.07 t_{\rm br,3} T_{\rm br,5}^{2.10} \, \rm days,
        \end{aligned}
    \end{split}
\end{equation}
and
\begin{equation}
    \label{eq: t_transparency}
    t_{tr}=19.5 \sqrt{M_{\rm env,0}\kappa_{0.34}v_{\rm s*,8.5}^{-1}}.
\end{equation}
The lower limits are set by the shock breakout time and the condition that light-travel time effects be negligible - after roughtly $2 R / c \sim 1\,R_{12} \, {\rm min}$. The upper limits are set by the time at which the temperature drops to 0.8~eV, and the time at which the photon escape time from deep within the envelope becomes comparable to the dynamical time. Note the slight difference between BSG and RSG validity times (the latter valid from $3R/c$ to $t_{\rm 0.7 eV}$). In BSG's, recombination occurs on average at slightly higher temperature, and the assumption of LTE fails after $t_{0.8 \, \rm eV}$ (see \S~\ref{sec: NLTE}).

\subsection{Frequency Dependent Formula}
\label{sec: freq dept formula}
Our frequency dependent model is derived piecwise, in the same way as in \citetalias{morag_shock_2024}, based on the strength of the absorption opacity $\kappa_{\rm abs,\nu}$ relative to the scattering opacity $\kappa_{\rm es}$. At frequencies where $\kappa_{\rm abs,\nu}>\kappa_{\rm es}$, the emitted flux may be approximated as a blackbody with a frequency-dependent thermal depth (surface of last absorption) $r_{\rm col,\nu}$, and corresponding frequency-dependent color temperature $T_{\rm col,\nu}=T(r_{\rm col,\nu})$, given by
\begin{equation}
    L_{\nu,\rm BB}=4\pi r_{\rm col,\nu}^2 B_{\nu}(T_{\rm col,\nu}).
    \label{eq: freq dept blackbody Tcol rcol prescription}
\end{equation}
At frequencies where the absorption opacity is smaller than the scattering opacity, $\kappa_{\rm abs,\nu}<\kappa_{\rm es}$, we base our approximation on the flux $f_\nu$ emitted by a semi-infinite planar slab of temperature $T$ in the two-stream approximation \citep{rybicki_radiative_1979}, arriving at 
\begin{equation}
    L_{\nu,\epsilon}=\frac{\left(4\pi\right)^{2}}{\sqrt{3}}r_{col,\nu}^{2}\frac{\sqrt{\epsilon_{\nu}}}{1+\sqrt{\epsilon_{\nu}}}B_{\nu}(T_{col,\nu}),
    \label{eq: epsilon prescription}
\end{equation}
where
\begin{equation}
    \epsilon_\nu=\kappa_{\rm abs,\nu}/(\kappa_{\rm abs,\nu}+\kappa_{\rm es}).
    \label{eq: epsilon def}
\end{equation}

We first derive an expression describing the emission at the regime of relatively low absorption opacity, $\kappa_{\rm abs,\nu}<\kappa_{\rm es}$, which occurs primarily at intermediate frequencies near and below the Planck peak. Absorption in this regime is dominated by free-free transitions with a small bound-free contribution. Neglecting the bound-free contribution, we approximate equation~(\ref{eq: Thermal Depth Integral k_abs_nu}), that defines the frequency-dependent thermal depth, as \citepalias[see also,][]{shussman_type_2016}
\begin{equation}
\int_{r_{\rm col,\nu}}^{\infty}\rho\sqrt{3\kappa_{\rm ff,\nu}\kappa_{es}}dr' =1.
    \label{eq: Thermal depth integral kff limit}
\end{equation}
Here we have neglected $\kappa_{\rm abs,\nu}$ with respect to $\kappa_{\rm es}$ and used
\begin{multline}
    \kappa_{\rm abs,\nu}\to \kappa_{\rm ff,\nu}=\\
    4.13\times10^{-31}g_{\rm ff}\rho T^{-1/2}(h\nu)^{-3}\left(1-\exp\left(-h\nu/T\right)\right) \, \rm cm^2 \, g^{-1},
\end{multline}
where the density $\rho$ is in cgs, temperature $T$ is in ergs. We approximate the gaunt factor $g_{\rm ff}\sim0.7\left(h\nu/T\right)^{-0.27}$.

Solving equation~(\ref{eq: Thermal depth integral kff limit}) using the analytic \citetalias{rabinak_early_2011}/\citetalias{sapir_uv/optical_2017} spherical phase density profiles (equations \ref{eq:rho_rt_vsstar} - \ref{eq:T_rt_bo}) we obtain the radius, temperature and opacity at the thermal depth,
\begin{equation}
    r_{\rm col,\nu} = 1.32 \times 10^{14} R_{13}^{-0.01}
    (f_{\rho}M_0)^{0.11} {\rm v_{\rm s*,8.5}^{0.74}}
    \kappa_{0.34}^{0.04} t_{\rm d}^{0.77} \nu_{eV}^{-0.09} \, \rm cm,
\end{equation}
\begin{equation}
     T_{\rm col,\nu} = 2.13 \, R_{13}^{0.28}
    {\rm v_{s*,8.5}^{0.16}} \nu_{\rm eV}^{0.25} \kappa_{0.34}^{-0.06}
    t_{\rm d}^{-0.45} \, \rm eV,
\end{equation}
\begin{equation}
    \kappa_{\rm ff,\nu} =
    0.02 \, (f_\rho M_0)^{-0.07}
    R_{13}^{-0.22} {\rm v_{\rm s*,8.5}^{-0.63}}
    \kappa_{0.34}^{-0.30}
    t_{\rm d}^{-0.14}
    \nu_{\rm eV}^{-1.66} \,
    \rm cm^2 \, g^{-1}.
\end{equation}
We find that modifying the expression for $r_{\rm col,\nu}$ to
\begin{equation}
     r_{\rm col,\nu} = R + 1.32 \times 10^{14} R_{13}^{-0.01}
    (f_{\rho}M_0)^{0.10} {\rm v_{\rm s*,8.5}^{0.74}}
    \kappa_{0.34}^{0.04} t_{\rm d}^{0.77} \nu_{eV}^{-0.09} \, \rm cm
\end{equation}
while keeping the expressions for $T_{\rm col,\nu}$ and $\kappa_{\rm ff,\nu}$ unchanged provides a good description of the spectrum also at the planar phase (and at the transition from planar to spherical evolution). 
In "break notation" we have
\begin{equation}
    r_{\rm col,\nu} = R + 2.02 \times 10^{13} L_{\rm br,42.5}^{0.48} T_{\rm br,5}^{-1.97}
    \kappa_{0.34}^{-0.08} \tilde{t}^{0.77} \nu_{\rm eV}^{-0.09} \, \rm cm,
    \label{eq: r_col_nu_br_notation}
\end{equation}
\begin{equation}
    T_{\rm col,\nu} = 5.71 \, L_{\rm br,42.5}^{0.06} T_{\rm br,5}^{0.91}
    \kappa_{0.34}^{0.22}
    \tilde{t}^{-0.45} \nu_{\rm eV}^{0.25} \, \rm eV,
    \label{eq: T_col_nu_br_notation}
\end{equation}
\begin{equation}
    \kappa_{\rm ff} = 0.03 \, L_{\rm br,42.5}^{-0.37} T_{\rm br,5}^{0.56} 
    \kappa_{0.34}^{-0.46}
    \tilde{t}^{-0.14}
    \nu_{\rm eV}^{-1.66} \, \rm cm^2 \, g^{-1}.
    \label{eq: k_col_nu_br_notation}
\end{equation}

The thermal depth values can then be inserted into equation (\ref{eq: epsilon prescription}) with $\epsilon_\nu=\kappa_{\rm ff,\nu}/(\kappa_{\rm ff,\nu}+\kappa_{\rm es})$ in order to describe the emission in the low absorption frequency range.

For frequency regions with strong absorption, where $\kappa_{\rm abs,\nu}>\kappa_{\rm es}$, we return to equation (\ref{eq: freq dept blackbody Tcol rcol prescription}). The thermal depth at these frequencies is located at the outer edge of the ejecta, where the density decreases sharply and the temperature, determined by the escaping photons, is nearly uniform. We therefore approximate $r_{\rm col,\nu}\approx const. (\nu)$ and $T_{\rm col,\nu}\approx const. (\nu)$ for these frequencies, and describe the emission as a gray blackbody $L_{BB}$, equation (\ref{eq:L_nu_BB_formula}). At low frequencies where the free-free opacity dominates, we find numerically that the luminosity is well approximated by $L_{BB}(0.85 \, T_{\rm col})$. Meanwhile, at frequencies near and above the Planck peak, where atomic transitions dominate, we use both the simulations and a separate analytic estimate (see \S~\ref{sec: expansion opac}) to improve upon the approximate $L_{\rm BB} (0.74 \, T_{\rm col})$ description of the UV suppression of \citetalias{morag_shock_2023}, replacing the suppression factor $0.74$ with a function of $(R,t)$ lying in the range $[0.6,1]$.

The combined freq-dept formula is thus
\begin{equation}
\label{eq:Lnu epsilon final}
L_{\nu} = \begin{cases}
\left[L_{\rm BB} (0.85 \, T_{\rm col})^{-m} + L_{\nu,\epsilon}^{-m}\right]^{-1/m} & h\nu<3.5 T_{\rm col} \\
1.2 \times L_{ \rm BB}(0.85 R_{13}^{0.13} t_{\rm d}^{-0.13} \times T_{\rm col}) & h\nu>3.5 T_{\rm col},
\end{cases}
\end{equation}
where $m=5$, and $L_{\nu,\epsilon}$ is again given by equations (\ref{eq: epsilon prescription}) and (\ref{eq: epsilon def}) with the choice $\kappa_{\rm abs,nu} \to \kappa_{\rm ff,\nu}$. The 1.2 factor accounts for modest UV excess we observe in our results at the planck peak due to the presence of strong lines. The frequency slope in the Raleigh-Jeans regime is similar, but slightly lower than the blackbody value $L\sim\nu^2$.

equation (\ref{eq:Lnu epsilon final}) can be further simplified to be given in terms of only $L$ and $T_{\rm col}$, with moderate decrease in the approximation's accuracy, 
\begin{multline}
     L_{\nu} = \\
   \begin{cases} \frac{\pi}{\sigma}\frac{L}{T_{\rm col}^4} \left[\, \left(\frac{B_{\nu}(0.85 T_{\rm col})}{(0.85)^4} \right)^{-m} + \right.\\
        \left. \left( \frac{8}{\sqrt{3}} x^{-0.18} T_{\rm col,5}^{-0.12} \frac{\sqrt{\epsilon_{\rm a}}}{1+\sqrt{\epsilon_{\rm a}}} B_\nu(1.71 \, x^{0.25} T_{\rm col}) \right)^{-m} \right]^{-1/m} & h\nu<3.5 T_{\rm col} \\
        \quad & \quad \\
        1.2\times L_{\rm BB}(1.11 L_{42.5}^{0.03} T_{5}^{0.18} \times T_{\rm col}) & h\nu>3.5T_{\rm col},
   \end{cases}
   \label{eq:Lnu epsilon simplified}
\end{multline}
where $x=h\nu/T_{\rm col}$, $T_{\rm col} = 5 \, T_{\rm col,5} \, \rm eV$, and $\epsilon_{\rm a} = 0.0051 \, x^{-1.66} T_{\rm col,5}^{-1.10}$. $L$ and $T_{\rm col}$ are given by equations (\ref{eq:L_gray_br}) and (\ref{eq:Tcol_gray_br}).

\section{Numeric Results}
\label{sec: numeric results}
\begin{figure}
    \includegraphics[width = \columnwidth]{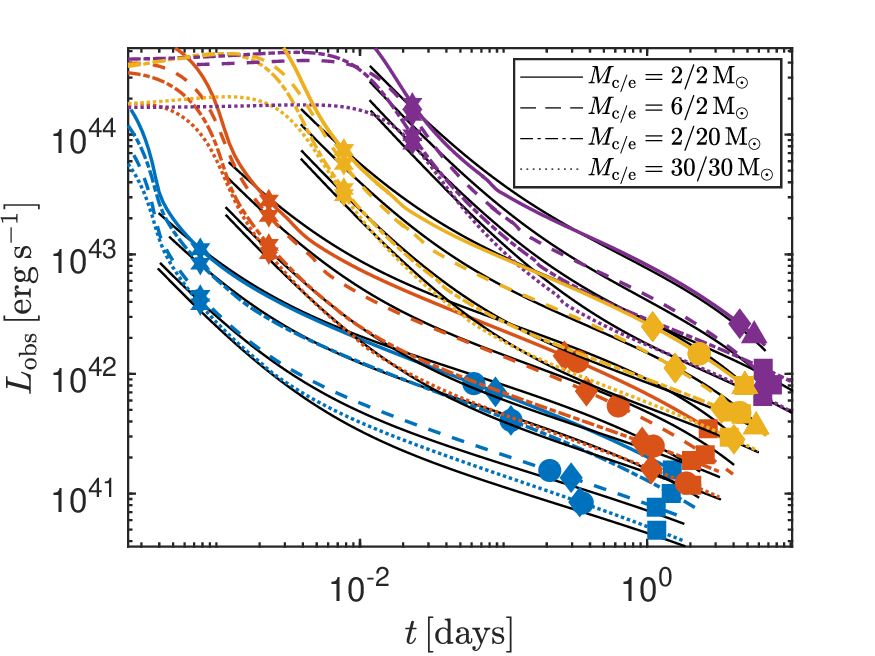}
    \caption[LMSW LTT caption]{Numerically derived bolometric luminosities, shown in color for different radii ($R_{13}$=0.1, 0.3, 1, 3 - blue to purple), mass combinations ($M_{c/e}$ denote core and envelope masses) and explosion energy of $E=10^{51}$ erg, compared to our analytic results, equation~(\ref{eq:L_gray_br}) in black. The numeric results shown include light-travel time effects,
    which modify $L$ at $t\lesssim R/c$. The symbols indicate the various validity times: $\star$ corresponds to $2R/c$ ; $\blacklozenge$ to the homologous time (equation 16, \citetalias{morag_shock_2023}); $\bullet$ to $t_{\rm ph}$ (equation 17, \citetalias{morag_shock_2023});  $\blacktriangle$ to $t_{\rm tr}/a$ (equation \ref{eq: t_transparency}); $\blacksquare$ to $t_{\rm 0.8 eV}$ (equation \ref{eq: t_08eV}).}
    \label{fig:L_RMser}
\end{figure}

\begin{figure}
    \includegraphics[width = \columnwidth]{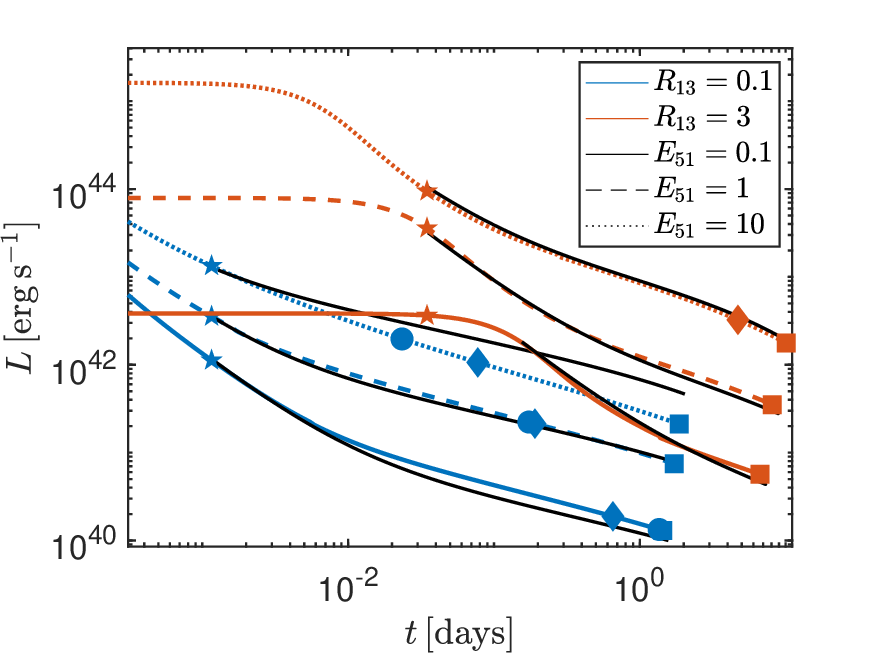}
    \caption{Numerically derived bolometric luminosities (in color) for several progenitor radii and explosion energies, compared to our gray formula (equation \ref{eq:L_gray_br}) in black, ignoring light-travel time effects. The core and envelope masses in all calculations shown are $M_{\rm c}=M_{\rm e}=10 M_{\odot}$. The case of $R_{13}=0.1$, $E_{51}=10$ shows a larger than average deviation from the formula in the spherical phase, in agreement with MG results (see text). The meaning of the symbols is the same as in fig.~\ref{fig:L_RMser}. For $R_{13}=3$, $E_{51}=0.1$, the early validity time is determined by the shock crossing time $t_{\rm bo}$, which is longer than $3 R/c$.}
    \label{fig:L_REser}
\end{figure}

\begin{figure}
    \includegraphics[width = \columnwidth]{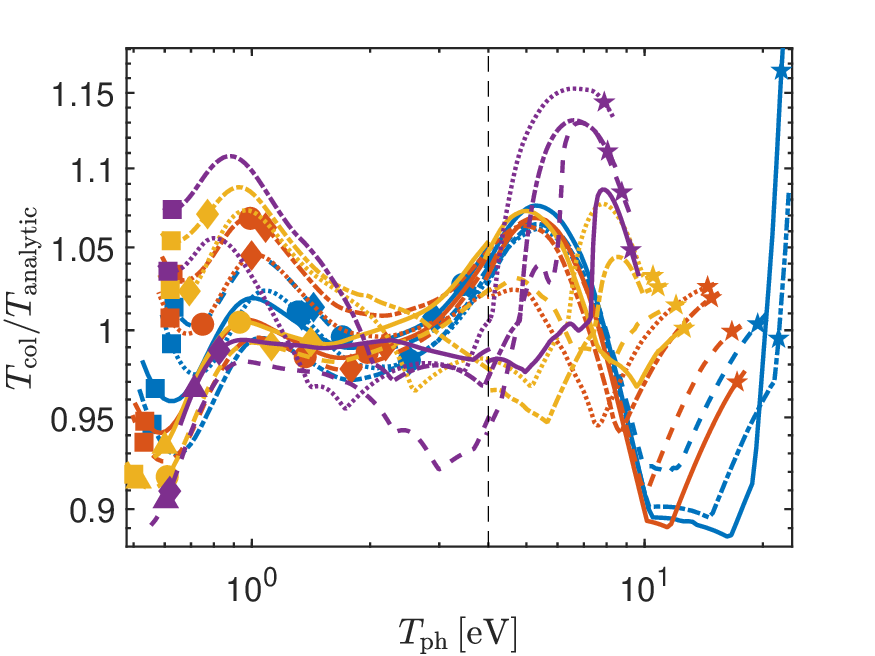}
    \caption{The ratio of numerically derived color temperatures to our analytic result, equation~(\ref{eq:Tcol_gray_br}), for different progenitor radii and mass combinations (corresponding to same styles and colors as in fig. \ref{fig:L_RMser}) and explosion energy of $E=10^{51} \rm \, erg$, plotted as a function of the photospheric temperature given by equation 4 in \citetalias{sapir_uv/optical_2017}. Line colors and styles and validity time marks are the same as in Fig.~\ref{fig:L_RMser}. The vertical dashed line denotes the 4~eV transition between our opacity table and the TOPS table (see text). The meaning of the symbols is the same as in fig.~\ref{fig:L_RMser}.}
    \label{fig:Tcol_RMser}
\end{figure}

\begin{figure}
    \includegraphics[width = \columnwidth]{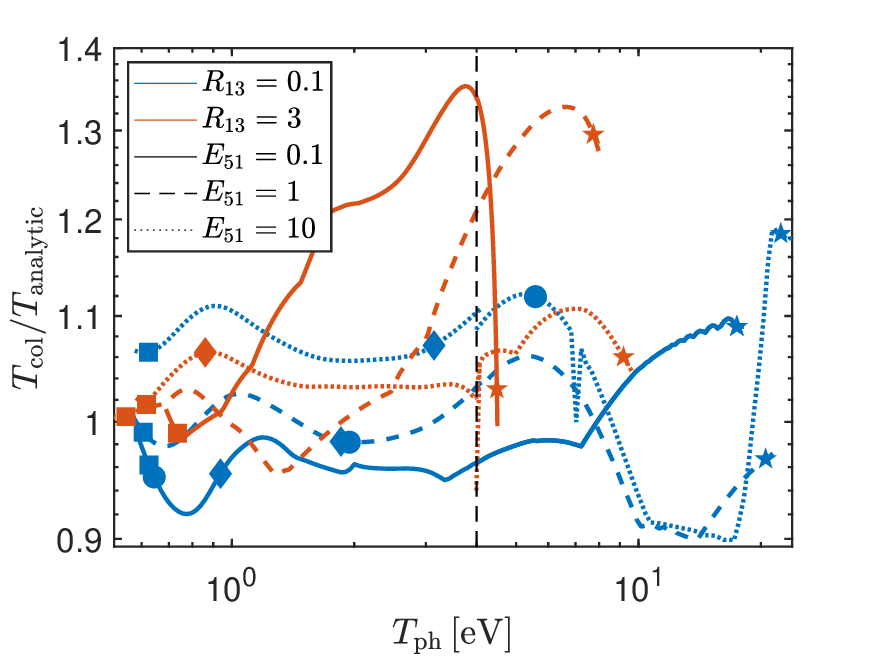}
    \caption{The ratio of numerically derived color temperatures for various explosion energies and radii. The numeric results for $R_{13}=3$, $E_{51}=0.1,1$ exhibit large (up to $\sim30\%$) deviations from the analytic formula, but the deviations of the results of MG simulations from the formula are smaller such that the approximation of equation \ref{eq:Itais_Prescription} works well for nearly all cases. The meaning of the symbols is the same as in fig.~\ref{fig:L_RMser}.}
    \label{fig:Tcol_REser}
\end{figure}

Figures \ref{fig:L_RMser}-\ref{fig:Tcol_REser} compare our numeric gray simulation results with our gray analytic solution, equations~(\ref{eq:L_gray_br}, \ref{eq:Tcol_gray_br}). The breakout parameters $\rho_{\rm bo}$ and $\beta_{\rm bo}={\rm v}_{\rm bo}/c$ are extracted from the simulation by fitting the luminosity of the breakout pulse to the table of \citetalias{sapir_non-relativistic_2011}, and these, along with the known $R$, are used to determine the physical parameters $v_{\rm s,*}$,\,$f_\rho M$ used in the equations without additional fitting. The RMS agreement is roughly $\sim10\%$ and $\sim5\%$ for $L(t)$ and $T_{\rm col}(t)$ respectively\footnote{The case $R_{13}=0.1,\, E_{51}=10$, which exhibits somewhat larger deviation in luminosity (fig. \ref{fig:L_REser}) is excluded from our analysis for $L$ and is considered somewhat outside our validity range. Likewise, the cases $R_{13}=3$,\,$E_{51}=(0.1,1)$ exhibit up to $\sim 30\%$ deviation, though the latter is not reproduced in the multigroup simulations.}. Fig. \ref{fig:L_RMser}, showing varying core and envelope masses combinations, also includes light-travel time effects, which are important during shock breakout and act to smear the observed breakout pulse (see footnote in \citetalias{morag_shock_2023} describing how they are calculated). Similarly to the case of RSG's, the light travel-time effects mark the early validity time, $t\sim2 R/c$, beyond which the effects are not important. Fig. \ref{fig:L_REser}, which shows the effect of varying explosion energy, does not include these effects. We also observe that varying the core radius between $R_{\rm core}/R\in[10^{-3},0.03]$, does not change the agreement with our formulae.

\begin{figure*}
    \centering
    \includegraphics[width = \textwidth]{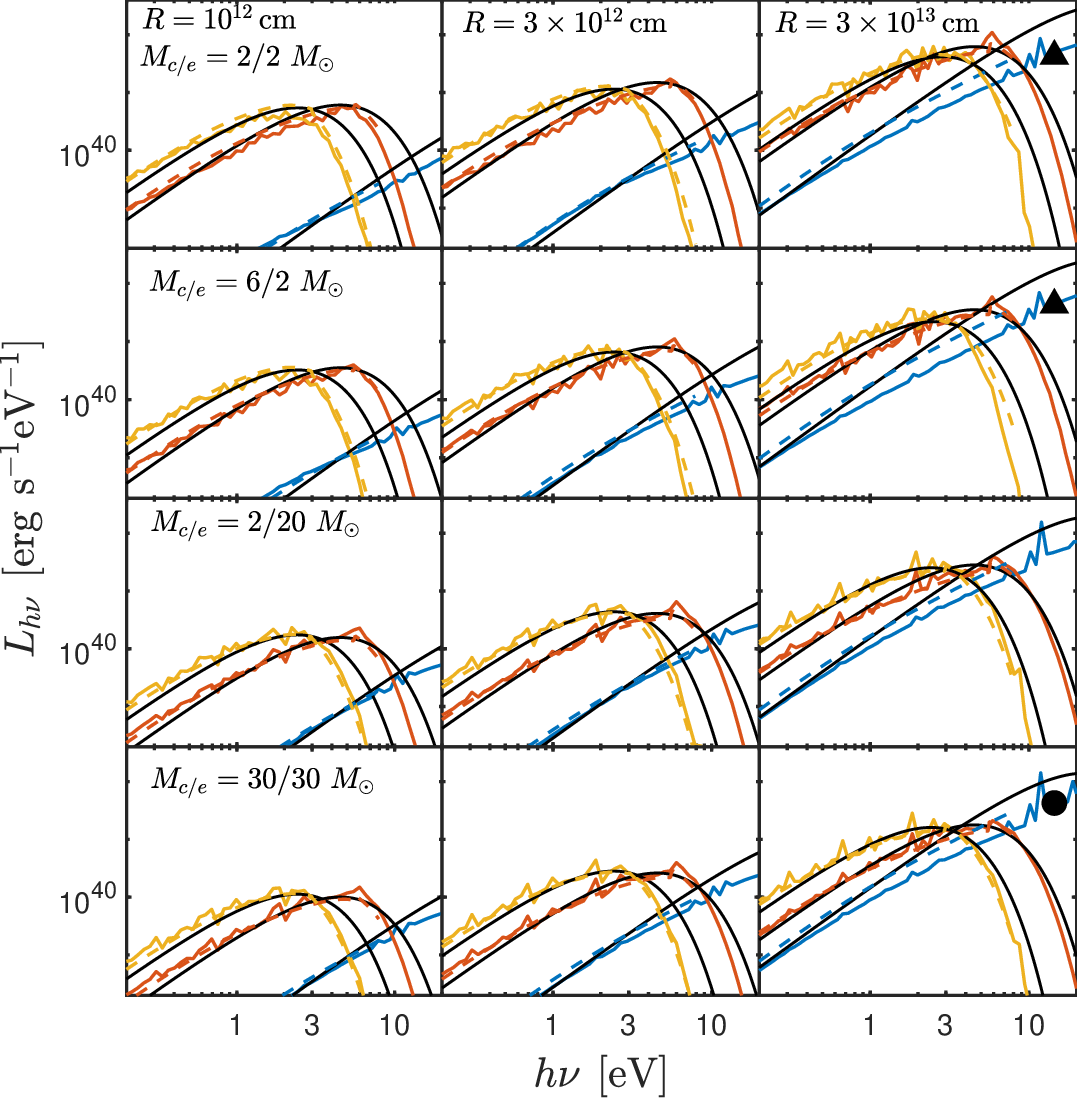}
    \caption{A comparison of our numeric (colored solid lines) and analytic results for the spectral luminosity, $\partial L/\partial h\nu$, for a range of progenitor radii (columns) and core/envelope mass combinations (rows). The black curves show our analytic gray approximation, equations~\ref{eq:L_gray_br} and \ref{eq:Tcol_gray_br}, and the dashed colored lines show the analytic frequency-dependent approximation, equation \ref{eq:Lnu epsilon final}, which is in excellent agreement with the simulation results. Blue/red/yellow colors denote results at times $2R/c$, $t_{1.5 \rm eV}$, $t_{0.8 \rm eV}$ respectively. The $\blacktriangle$ symbols mark cases for which the transparency time, $t_{\rm tr}/a$ is plotted instead of $t_{\rm 0.8 \, eV}$ in yellow, since the former is earlier than the latter (and hence the former determines the late validity time of our approximations). The $\bullet$ symbols mark cases where the shock-breakout time is larger than $2R/c$ (and hence determines the early validity time of our approximations).  
    }
    \label{fig:Lnu_tiles_RMser}
\end{figure*}

\begin{figure*}
    \centering
    \includegraphics[width = \textwidth]{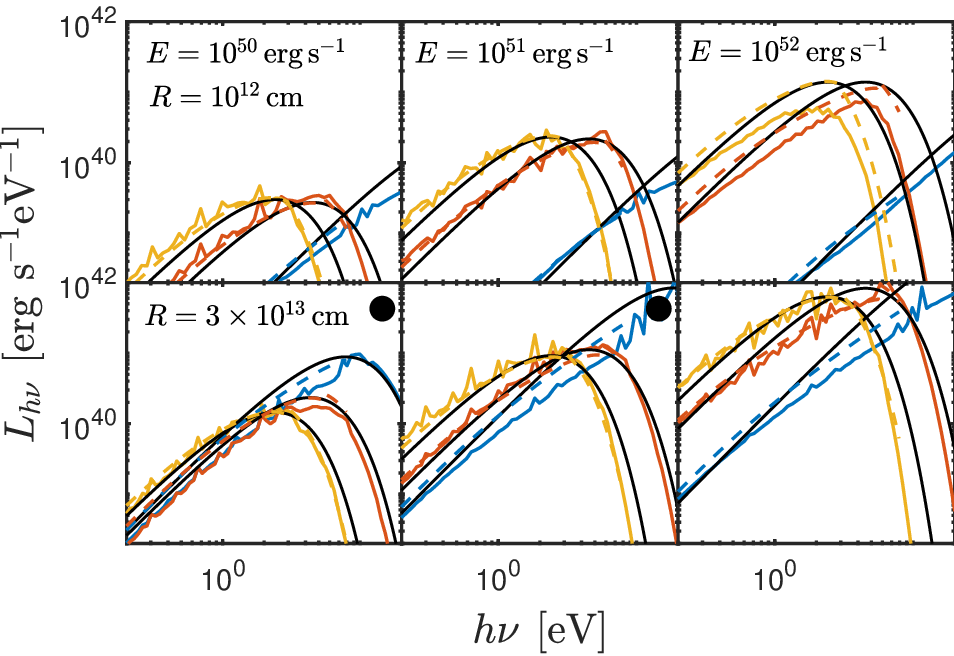}
    \caption{A comparison of our numeric (colored solid lines) and analytic results for the spectral luminosity, $\partial L/\partial h\nu$, for a range of progenitor radii (rows) and explosion energies (columns) with $M_{\rm core}=M_{\rm envelope}=20 M_{\odot}$. Line types and symbols are the same as in fig.~\ref{fig:Lnu_tiles_RMser}. The case $R_{13}=0.1$,$E_{51}=10$ exhibits larger deviations of the analytic approximation from the simulations' results (see also fig. \ref{fig:L_REser}), and is considered to be outside the validity range of our approximations.}
    \label{fig:Lnu_tiles_REser}
\end{figure*}

In Figures \ref{fig:Lnu_tiles_RMser}-\ref{fig:Lnu_tiles_REser} we compare the SED results from multigroup numeric simulations to our gray (equations~\ref{eq:L_gray_br}-\ref{eq:Tcol_gray_br}) and frequency-dependent formulae (equation~\ref{eq: epsilon prescription}). In agreement with our previous works, the SED is approximately a blackbody, with minor 10 percent deviations in the Raleigh Jeans tail and prominent suppression in the ultraviolet. We find 20-35\% RMS agreement between the frequency-dependent formulae and simulations, excluding for the $R=10^{12}$ cm, $E=10^{51}$ erg calculation. The slightly simplified frequency-dependent formula (equation~\ref{eq:Lnu epsilon simplified}) yields an RMS inaccuracy of 35-40\%. Both of the uncertainties quoted above include (in a sum of squares sense) the uncertainties in our numeric model, based on a comparison to a different method of estimating the SED  as described in sec.~\ref{sec: expansion opac}.

\section{Expansion Opacity and Local Thermal Equillibrium}
\subsection{Expansion Opacity and Finite Frequency Resolution}
\label{sec: expansion opac}
We show here, as we did in \citetalias{morag_shock_2024}, that our numeric code should correctly describe the approximate SED despite being coarse in frequency resolution relative to transition line widths and despite not including `expansion opacity' effects \citep[see e.g.][]{friend_stellar_1983,Eastman_Spectrum_1993,castor_radiation_2007,rabinak_early_2011}. 

In our simulations, for each frequency bin we use the Rosseland mean of the opacity for the diffusion calculation and the frequency averaged opacity for the emission/absorption calculation. In the presence of large velocity gradients, such that the Doppler shift of the plasma-frame photon frequency across a spatial resolution element, $\Delta r$, is comparable to the frequency separation between strong lines, the effective photon mean-free-path may be much smaller than that derived from the Rosseland mean, 
\begin{equation}\label{eq:l_exp}
  l_{\rm exp}\approx\frac{c}{\rm v}\frac{\Delta\nu}{\nu}r,
\end{equation}
where $\Delta\nu/\nu$ is the frequency difference between adjacent `strong' lines (with optical depth $\tau$>1 taking into account Doppler shift, see \citetalias{morag_shock_2024} and references therein). The ratio $l_{\rm Ross}/l_{\rm exp}$, determines when expansion opacity effects may be significant.

\begin{figure}
    \centering
    \includegraphics[width=\columnwidth]{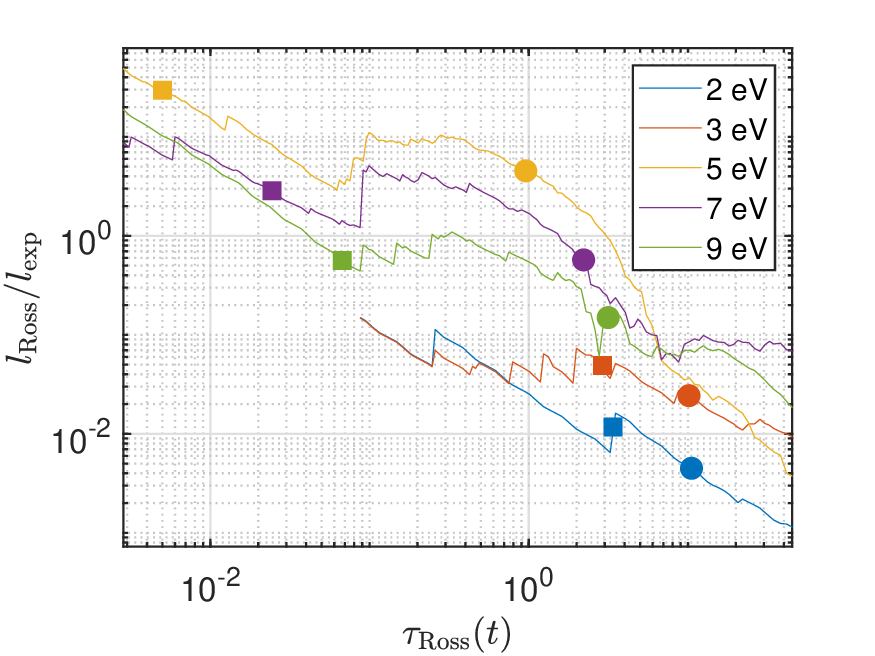}
    \includegraphics[width=\columnwidth]{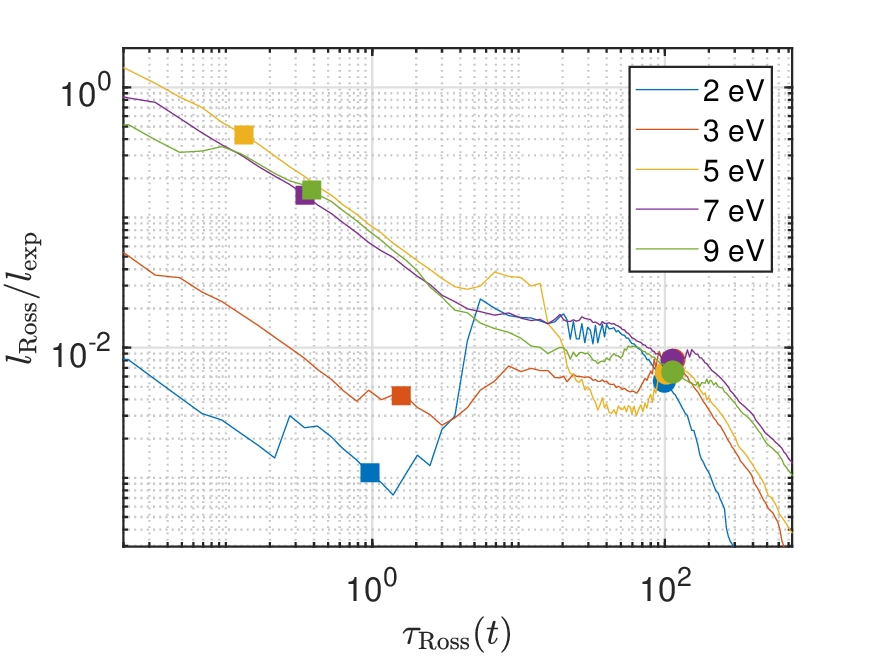}
    \caption{Rosseland mean-free-path relative to expansion mean-free-path at different frequency bins as a function of binned Rosseland mean optical depth $\tau_{\rm Ross}$. $\bullet$ and $\blacksquare$ symbols indicate the location of the diffusion depth and thermal depth (estimated without expansion opacity) respectively. Results are shown at the end of our validity time, $t_{0.8 \, \rm eV}$ (around recombination) when the effect of expansion opacity is strongest, for two different parameter choices: Top, $R_{13}=1, \, M_{\rm c}=M_{\rm e}=2 M_\odot, \rm E_{51}=1$; Bottom, $R_{13}=3, \, M_{\rm c}=M_{\rm e}=10 M_\odot, \rm E_{51}=10$. These examples are representative of those obtained for our examined ranges of radii, masses, and explosion energies.}
    \label{fig:lRoss_lDopp}
\end{figure}

Using our high-frequency opacity table, we extract the locations of strong lines and examine the above ratio at the latest validity time in our analysis, $t_{0.8 \, \rm eV}$, when expansion effects are strongest. Similarly to our analysis of RSG's, we find that the Rosseland mean dominates the opacity for most frequencies. At frequencies above the Planck peak (5-7 eV), the velocity gradient effect becomes significant at late time at the outer edge of the ejecta. At these frequencies, $l_{\rm Ross} \sim l_{\rm exp}$ at the diffusion depth or further out, see~fig. \ref{fig:lRoss_lDopp}. However, since the temperature does not vary rapidly beyond this radius, the impact of the modification of the mean-free-path, compared to the Rosseland mean, is not expected to be large, as we demonstrate below.

\begin{figure}
    \centering
    \includegraphics[width=\columnwidth]{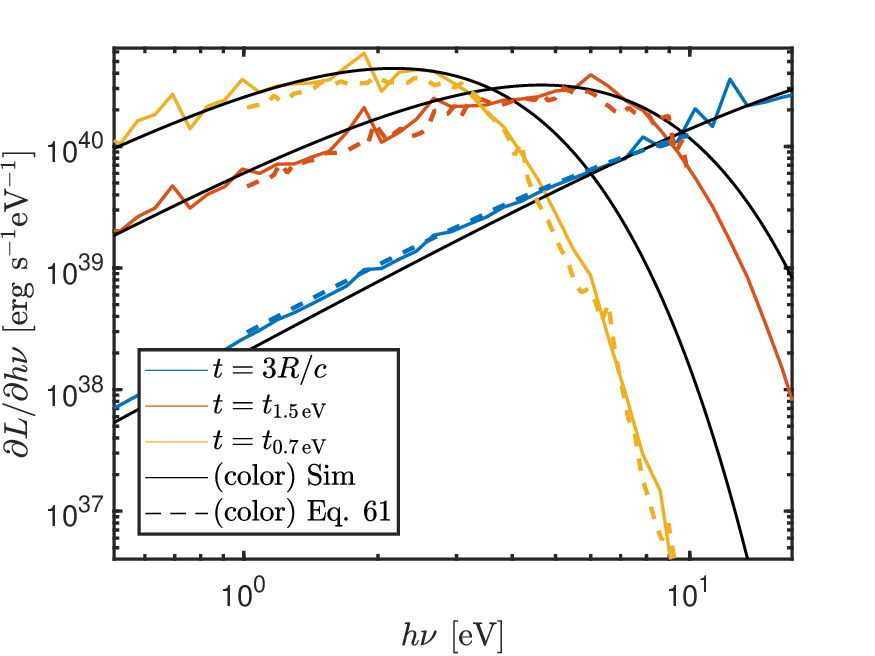}
    \includegraphics[width=\columnwidth]{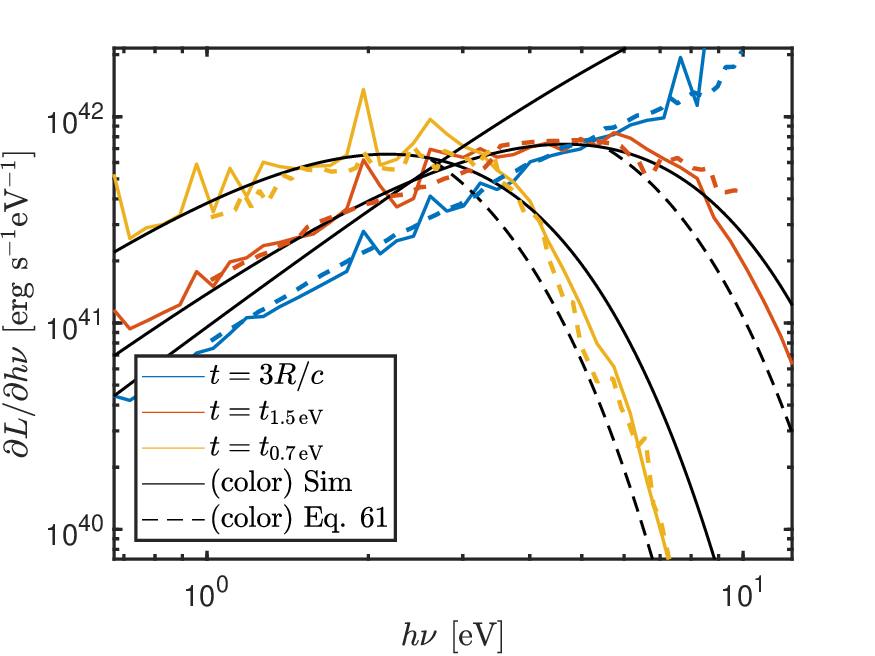}
    \caption{A comparison of the flux obtained in the numeric simulations, that do not include the effects of expansion opacity, with the flux obtained using the analytic approximations of equations~(\ref{eq: freq dept blackbody Tcol rcol prescription}) and~(\ref{eq: epsilon prescription}) with $r_{\rm col,\nu}$ (equation~\ref{eq: Thermal Depth Integral k_abs_nu}) calculated with the density and temperature profiles obtained in the simulation, but with a high resolution, $\Delta\nu/\nu\sim10^{-5}$ opacity table and taking into account the Doppler shifts of the lines 
    (The gray approximation, equations (\ref{eq:L_gray_br})-(\ref{eq:Tcol_gray_br}) shown in black). The high-resolution spectrum has been averaged over frequency bins, without affecting the SED.
    Top (bottom): progenitor radius $R = 3\times10^{12}$ cm ($R=10^{14}$ cm), explosion energy $E=10^{51}$ erg, and core and envelope masses $M_{\rm env}=M_{\rm core}=10 M_{\odot}$.}
    \label{fig:Lnu_vs_thrmdpth_Doppler}
\end{figure}

As in \citetalias{morag_shock_2024}, to test the sensitivity of our numeric results to the finite frequency resolution used and to the effects of `expansion opacity', we perform a separate calculation of the SED in `post-processing', using the numeric plasma density and temperature profiles. The SED is estimated by calculating the frequency-dependent thermal depth $r_{\rm col,\nu}$ and corresponding color temperature $T_{\rm col,\nu}$ as determined by equation (\ref{eq: Thermal Depth Integral k_abs_nu}), using the full high-resolution-frequency opacity table and including the effect of Doppler shifts (in this calculation we do not bin the photons into energy groups). The spectral luminosity in this test is then determined by
\begin{equation}
   L_{\nu,\rm Dopp} = 
   \begin{cases}
        \rm Eq. \, (\ref{eq: freq dept blackbody Tcol rcol prescription}) & \tau_{\rm es} ( \tau_{*,\nu}=1) \leq f_{\rm cut} \\
        \rm Eq. \, (\ref{eq: epsilon prescription}) & \tau_{\rm es} ( \tau_{*,\nu}=1) > f_{\rm cut},
   \end{cases} 
   \label{eq: Doppler Integral}
\end{equation}
where the SED is not sensitive to the value of $f_{\rm cut}$, chosen in the range 0.1-3.
The SED obtained agrees with that obtained directly from the numeric simulations to $\sim 10-20 \%$ (see examples in fig. \ref{fig:Lnu_vs_thrmdpth_Doppler}), implying that our numeric SED results are accurate at this level. 

\subsection{Deviations from LTE Ionization and Excitation}
\label{sec: NLTE}
The opacity tables that we use in our numeric calculations were constructed assuming LTE in ionization and excitation states. We show below that the deviations from LTE are not expected to be large, and hence are not exepcted to affect the SED significantly, based on an analysis of the relevant interaction rates (similar to the analysis of \citetalias{morag_shock_2024}). We note that one of the reasons for this result is that during shock-cooling, the photons dominate the energy density in the ejecta and are nearly thermally distributed (see, e.g., fig. 9 of \citetalias{morag_shock_2024}). This is in contrast to the later nebular phase, where the photons do not dominate the energy density and are far from a Planck distributed. 

\begin{figure}
    \centering
    \includegraphics[width=\columnwidth]{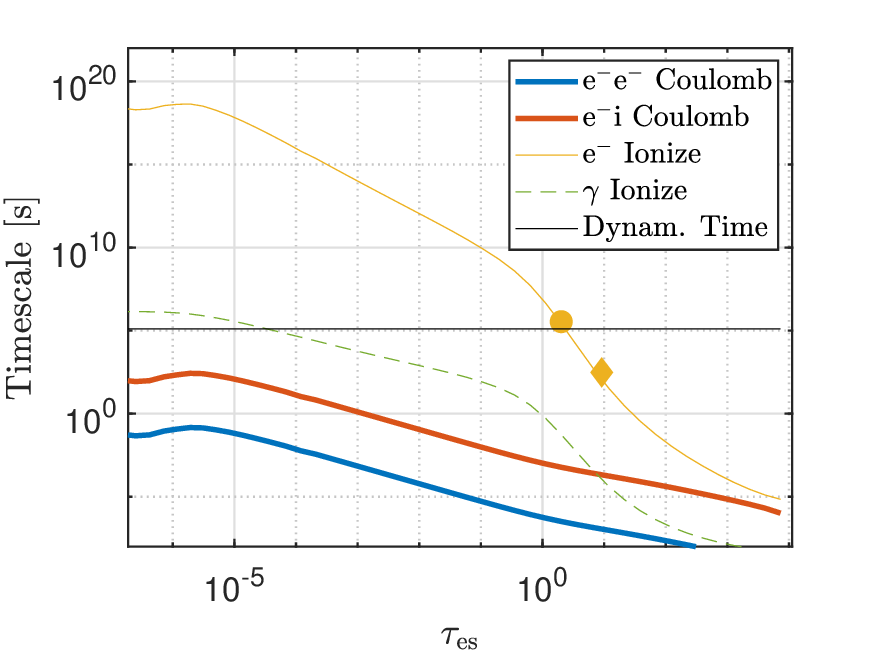}
    \includegraphics[width=\columnwidth]{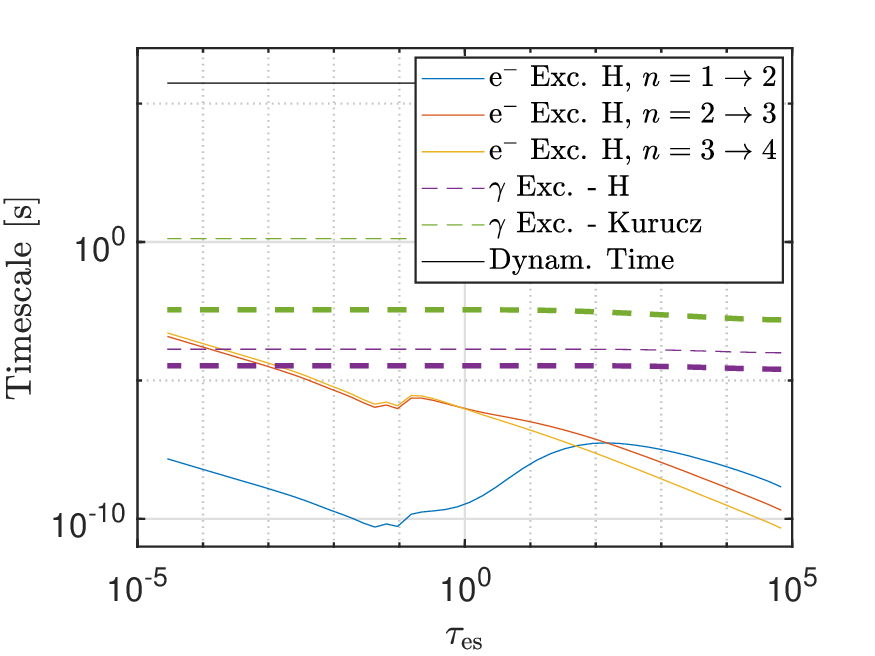}
    \caption{Example inverse interaction rates as a function of scattering optical depth $\tau_{\rm es}$ shown at recombination time, $t_{0.8 \rm \, eV}$, when they are slowest compared to the (inverse of the) dynamical time. Collisional electron processes and photon processes are shown in solid  and dashed lines respectively. At top, time scales for ionization and for velocity distribution thermalization by Coulomb interactions are shown for $R_{13}=0.1,\rm M_{\rm c}=M_{\rm e}=2M_\odot,\, E_{51}=1$. At bottom, electron collisional excitation and photoexcitation time scales are shown. Thin (thick) dashed lines denote maximum (average) photon excitation timescales based on an average of the atomic lines (see text) for $R_{13}=3,\rm M_{\rm c}=M_{\rm e}=30M_\odot,\, E_{51}=1$. The characteristic time scales of all processes, except electron collisional ionization, are much shorter than the dynamical time.}
    \label{fig:ion_exc_rates}
\end{figure}

Collision, excitation and ionization rates typical for the shock-cooling phase are shown in fig. \ref{fig:ion_exc_rates}. With the exception of electron impact ionization, the rates are large compared to the (inverse of the) dynamical time (including those of electron-electron and electron-ion Coulomb collisions, electron impact excitation, photo-excitation, and ionization). The relative low rate of impact ionization is not expected to lead to deviations from LTE ionization since ionization is dominated by photons and the energy distributions of both photons and electrons are close to those of LTE. 

\begin{figure}
\includegraphics[width=\columnwidth]{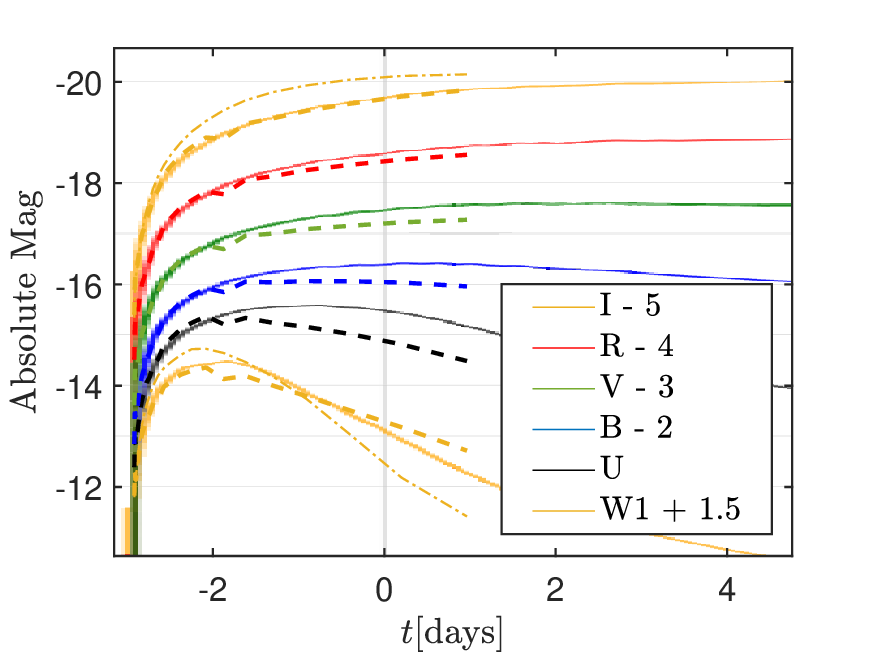}
\caption{A comparison of the \citet{singh_sn_2019} lightcurves obtained from a STELLA calculation (solid line) of a BSG progenitor explosion with those obtained by our calculations for similar progenitor and explosion parameters (we approximate the density profile as polytropic, see text). $t=0$ corresponds to peak light, magnitudes are in the VEGA system. Dashed lines and thin dash-dot lines show our results for opacity not including and including bound-bound transitions, respectively (the latter are shown only for two bands to reduce clutter). Better agreement is obtained when the bound-bound contribution is ignored, providing further evidence for the underestimation
of the effects of lines in STELLA calculations (see text).}
\label{fig: Singh-like}
\end{figure}

\section{Comparison to Previous Works}
\label{sec: STELLA compare}
In \citetalias{morag_shock_2024}, we compared our numeric SED results to those of several earlier numeric works that use the STELLA \citep{blinnikov_stella_2011} radiative transfer code to calculate the emission of radiation from shock breakout and cooling from stars with density profiles derived using the MESA \citep{paxton_modules_2015} stellar evolution code. We showed that our results, with progenitor density profiles approximated as simple polytropes, are in good agreement with those earlier results, with the important exception of the flux at the thermal-peak frequencies where strong lines are present. This demonstrates the validity of the diffusion approximation we use for the radiative transfer, and reconfirms the result \citepalias[see][]{sapir_uv/optical_2017} that the emission is not sensitive to deviations of the density profiles from polytropic profiles. It also suggests that STELLA calculations underestimate the effect of lines relative to us (see \citetalias{morag_shock_2024} for a discussion).

Here we carry out a similar comparison, reproducing STELLA lightcurves of \citet{singh_sn_2019}, that describe shock-cooling following an explosion of a BSG progenitor. We approximate the progenitor density profile as a polytropic profile, with $R=50 R_{\odot}$ and core and envelope masses of $M_{\rm c}=M_{\rm e}=7 M_\odot$ (this mass ratio is chosen arbitrarily for lack of information). The explosion energy is $E=1.7\times10^{51}$ erg. Fig.~\ref{fig: Singh-like} shows moderate agreement between our results and those of \citet{singh_sn_2019} for the multiband lightcurves in the first hours and days following the explosion. A better agreement is obtained when we turn off the bound-bound opacity in our calculations. This provides further evidence for the underestimation of the effects of lines in STELLA calculations.

\section{Discussion and Summary}
\label{sec: conclusion}

In preceding papers of this series, \citetalias{morag_shock_2023} and \citetalias{morag_shock_2024}, we provided a simple analytic description of the time-dependent luminosity, $L$, and color temperature, $T_{\rm col}$, as well as of the small ($\simeq10\%$) deviations of the spectrum from blackbody at low frequencies, $h\nu< 3T_{\rm col}$, and of  `line dampening' at $h\nu> 3T_{\rm col}$, for explosions of RSGs with convective polytropic envelopes (without significant circum-stellar medium). Here, we extended our work to provide similar analytic formulae for explosions of BSGs with radiative polytropic envelopes. The approximations for $L(t)$ and $T_{\rm col}(t)$ are given in equations~(\ref{eq:L_gray_br}) and (\ref{eq:Tcol_gray_br}), and the frequency-dependent deviations from blackbody are given by equation~(\ref{eq:Lnu epsilon final}). A slightly less accurate approximation for the frequency-dependent deviations, that depends only on $L$ and $T_{\rm col}$, is given in equation~(\ref{eq:Lnu epsilon simplified}). The formulae describing our approximations for shock cooling emission for both RSGs and BSGs are summarized in the appendix. They are valid until significant recombination of Hydrogen.

The analytic formulae were calibrated against a large set of 1D `gray' and multi-group (frequency-dependent) calculations for a wide range of progenitor parameters (mass, radius, core/envelope mass ratios) and explosion energies using the opacity tables we constructed (and made publicly available), that include the contributions of bound-bound and bound-free transitions. They reproduce the numeric $L$ and $T_{\rm col}$ to within 10\% and 5\% accuracy, and the spectral energy distribution to within $\sim20-40\%$. We have shown \citepalias[][and \S~\ref{sec: expansion opac}, \S~\ref{sec: NLTE}]{morag_shock_2024} that the SED is not sensitive to the effects of expansion opacity and deviations from LTE ionization and excitation. 

Our numeric results are in good agreement \citepalias[][and \S~\ref{sec: STELLA compare}]{morag_shock_2023,morag_shock_2024} with those of STELLA calculations of shock cooling emission from the explosions of RSG and BSG progenitors with non-polytropic pre-explosion density profiles obtained from the MESA stellar evolution code. This demonstrates
the validity of the diffusion approximation we use for the radiative transfer, and reconfirms the result \citepalias[see][]{sapir_uv/optical_2017} that the emission is not sensitive to deviations of the density profiles from polytropic
profiles. We find that STELLA calculations underestimate
the effect of lines relative to our calculations \citepalias[see \S~\ref{sec: STELLA compare} and][for a discussion]{morag_shock_2024}.

\bibliographystyle{mnras}
\bibliography{references}

\appendix
\section{Summary of model equations}
\label{appendix}

We provide here the formulae describing shock cooling emission for both RSG's and BSG's.

\subsection{Gray Formulae}
The bolometric luminosity $L$ and the color temperature $T_{\rm col}$ (equations \ref{eq:L_gray_br}-\ref{eq:Tcol_gray_br} for BSG's), are given by
\begin{equation}
    L/L_{\rm br}=\tilde{t}^{-4/3}+\tilde{t}^{[-0.17 \, , \, -0.34]}\times A R_{13}^{[ 0\,,\,0.1]} \exp\left(-\left[at/t_{\rm tr}\right]^{\alpha}\right), \\
    \label{eq:L_trans_Appendix}
\end{equation}
\begin{equation}
    T_{\rm col}/T_{\rm col,br}=\min\left[0.97\,\tilde{t}^{-1/3},\tilde{t}^{[-0.45 \, , \, -0.40]}\right].
    \label{eq:T_trans_Appendix}
\end{equation}
Here, the $[ x, y]$ notation indicates values $x,y$ given for $n=3/2,3$ respectively, $\{A,a,\alpha\} = [\{0.9,2,0.5\},\{0.79,4.57,0.73\}]$, $\tilde{t}=t/t_{\rm br}$, and we define $t=0$ as the time at which the breakout flux peaks. 

For RSG's, the break parameters (with br subscript) are given as a function of progenitor radius $R$, ejecta velocity $\rm v_{\rm s*}$, and total ejecta mass $M$, by
\begin{equation}
\label{eq:t_br_of_vs_Appendix}
    t_{\rm br}= 0.86 \, R_{13}^{1.26} {\rm v_{\rm s*,8.5}^{-1.13}}
(f_{\rho}M_0\kappa_{0.34})^{-0.13}\,\text{hrs},
\end{equation}
\begin{equation}
\label{eq:L_br_of_vs_Appendix}
L_{\rm br}=3.69\times10^{42} \, R_{13}^{0.78} {\rm v_{\rm s*,8.5}^{2.11}}
(f_{\rho}M_0)^{0.11} \kappa_{0.34}^{-0.89}\,{\rm erg \, s^{-1}},
\end{equation}
\begin{equation}
\label{eq:T_br_of_vs_Appendix}
T_{\rm col,br}= 8.19 \, R_{13}^{-0.32} {\rm v_{\rm s*,8.5}^{0.58}}
(f_{\rho}M_0)^{0.03} \kappa_{0.34}^{-0.22}\,{\rm eV}.
\end{equation}
For BSG's (equations \ref{eq:t_br_of_vs}-\ref{eq:T_br_of_vs}) we have
\begin{equation}
\label{eq:t_br_of_vs_BSG_app}
    t_{\rm br}= 0.69 \, R_{13}^{1.32} v_{\rm s*,8.5}^{-1.16}
(f_{\rho}M_0\kappa_{0.34})^{-0.16}\,\text{hrs},
\end{equation}
\begin{equation}
\label{eq:L_br_of_vs_BSG_app}
L_{\rm br}=7.16\times10^{42} \, R_{13}^{0.55} v_{\rm s*,8.5}^{2.22}
(f_{\rho}M_0)^{0.22} \kappa_{0.34}^{-0.77}\,{\rm erg \, s^{-1}},
\end{equation}
\begin{equation}
\label{eq:T_br_of_vs_BSG_app}
T_{\rm col,br}= 9.49 \, R_{13}^{-0.38} {\rm  v_{\rm s*,8.5}^{0.62}}
(f_{\rho}M_0)^{0.07} \kappa_{0.34}^{-0.18}\,{\rm eV}.
\end{equation}
Here, $R= 10^{13}\,R_{13}$~cm, $\kappa=0.34 \, \kappa_{0.34} \, \rm \, cm^2 \, g^{-1}$, ${\rm v_{\rm s*}=v_{\rm s*,8.5}} \, 10^{8.5} \, \rm cm \,  s^{-1}$, $M_0$ denotes mass in units of solar mass, and $f_\rho\simeq1$ depends on the inner structure of the envelope (see equation~\ref{eq:rho_in}). $\rm v_{\rm s\ast}$ is related to the characteristic ejecta velocity $\rm v_\ast$ by (equation \ref{eq:vstar})
\begin{equation}
\label{eq:Avstar_Appendix}
    {\rm v_{\rm s\ast}}\approx 1.05 f_\rho^{-0.19}v_\ast,\quad {\rm v_\ast}\equiv\sqrt{E/M},
\end{equation}
where $E$ is the energy deposited in the ejecta.

\subsection{Frequency Dependent Formulae}
Our analytic approximation for the spectral luminosity of the shock cooling emission, taking into account deviations from a blackbody spectrum, is (equation \ref{eq:Lnu epsilon final})
\begin{equation}
\label{eq:Lnu epsilon final_Appendix}
L_{\nu} = \begin{cases}
\left[L_{\rm BB} (0.85 \, T_{\rm col})^{-m} + L_{\nu,\epsilon}^{-m}\right]^{-1/m} & h\nu<3.5 T_{\rm col} \\
1.2 \times L_{\rm BB}(0.85 R_{13}^{0.13} t_{\rm d}^{-0.13} \times T_{\rm col}) & h\nu>3.5 T_{\rm col},
\end{cases}
\end{equation}
with $m=5$ and (equations \ref{eq:L_nu_BB_formula}, \ref{eq: epsilon prescription})
\begin{equation}
    L_{\rm BB}=L\times\pi B_{\nu}(T_{\rm col})/\sigma T_{\rm col}^{4}
    \label{eq:L_nu_BB_formula_Appendix},
\end{equation}
\begin{equation}
 L_{\nu,\epsilon}=\frac{\left(4\pi\right)^{2}}{\sqrt{3}}r_{col,\nu}^{2}\frac{\sqrt{\epsilon_{\nu}}}{1+\sqrt{\epsilon_{\nu}}}B_{\nu}(T_{col,\nu}),\quad 
  \epsilon_\nu=\frac{\kappa_{\rm ff,\nu}}{\kappa_{\rm ff,\nu}+\kappa_{\rm es}}.  \label{eq: epsilon prescription _Appendix}
\end{equation}

For RSG's
\begin{equation}
    r_{\rm col,\nu} = R + 2.18 \times 10^{13} L_{\rm br,42.5}^{0.48} T_{\rm br,5}^{-1.97}
    \kappa_{0.34}^{-0.07} \tilde{t}^{0.80} \nu_{\rm eV}^{-0.08} \, \rm cm,
\end{equation}
\begin{equation}
    T_{\rm col,\nu} = 5.47 \, L_{\rm br,42.5}^{0.05} T_{\rm br,5}^{0.92}
    \kappa_{0.34}^{0.22}
    \tilde{t}^{-0.42} \nu_{\rm eV}^{0.25} \, \rm eV,
\end{equation}
\begin{equation}
    \kappa_{\rm ff} = 0.03 \, L_{\rm br,42.5}^{-0.37} T_{\rm br,5}^{0.56} 
    \kappa_{0.34}^{-0.47}
    \tilde{t}^{-0.19}
    \nu_{\rm eV}^{-1.66} \, \rm cm^2 \, g^{-1}.
\end{equation}
For BSG's the corresponding equations are (equations \ref{eq: r_col_nu_br_notation}-\ref{eq: T_col_nu_br_notation})
\begin{equation}
    r_{\rm col,\nu} = R + 2.02 \times 10^{13} L_{\rm br,42.5}^{0.48} T_{\rm br,5}^{-1.97}
    \kappa_{0.34}^{-0.08} \tilde{t}^{0.77} \nu_{\rm eV}^{-0.09} \, \rm cm,
\end{equation}
\begin{equation}
    T_{\rm col,\nu} = 5.71 \, L_{\rm br,42.5}^{0.06} T_{\rm br,5}^{0.91}
    \kappa_{0.34}^{0.22}
    \tilde{t}^{-0.45} \nu_{\rm eV}^{0.25} \, \rm eV,
\end{equation}
\begin{equation}
    \kappa_{\rm ff} = 0.03 \, L_{\rm br,42.5}^{-0.37} T_{\rm br,5}^{0.56} 
    \kappa_{0.34}^{-0.46}
    \tilde{t}^{-0.14}
    \nu_{\rm eV}^{-1.66} \, \rm cm^2 \, g^{-1},
\end{equation}
Here, $L_{\rm br}=L_{\rm br,42.5} 10^{42.5} \rm \, erg \, s^{-1}$, $T_{\rm col}=5 T_{\rm col,5}$ eV, and $\nu=\nu_{\rm eV}$ eV. 

$R$ is given in terms of the break parameters as
\begin{equation}
    R = 2.41\times10^{13} \, t_{\rm br,3}^{-0.1} \,L_{\rm br,42.5}^{0.55} \,T_{\rm br,5}^{-2.21} \,\text{cm}
\end{equation}
for RSGs, and as 
\begin{equation}
    R = 2.23\times10^{13} \, t_{\rm br,3}^{-0.1} \,L_{\rm br,42.5}^{0.56} \,T_{\rm br,5}^{-2.24} \,\text{cm}
\end{equation}
for BSG's (equation \ref{eq:R_br}).

A simpler approximation, that depends only on the $L$ and $T_{\rm col}$ and is slightly less accurate, is 
\begin{multline}
   L_{\nu} = \\
   \begin{cases} \frac{\pi}{\sigma}\frac{L}{T_{\rm col}^4} \left[\, \left(\frac{B_{\nu}(0.85 T_{\rm col})}{(0.85)^4} \right)^{-m} + \right.\\
        \left. \left( \frac{8}{\sqrt{3}} x^{-0.155} T_{5}^{-0.1} \frac{\sqrt{\epsilon_{\rm a}}}{1+\sqrt{\epsilon_{\rm a}}} B_\nu(1.63 \, x^{0.247} T_{\rm col}) \right)^{-m} \right]^{-1/m} & h\nu<3.5 T_{\rm col} \\
        \quad & \quad \\
        1.2\times L_{\nu,\rm BB}(1.11 L_{42.5}^{0.03} T_{5}^{0.18} \times T_{\rm col}) & h\nu>3.5T_{\rm col}
   \end{cases}
\end{multline}
for RSG's, and
\begin{multline}
   L_{\nu} = \\
   \begin{cases} \frac{\pi}{\sigma}\frac{L}{T_{\rm col}^4} \left[\, \left(\frac{B_{\nu}(0.85 T_{\rm col})}{(0.85)^4} \right)^{-m} + \right.\\
        \left. 
        \left( \frac{8}{\sqrt{3}} x^{-0.18} T_{\rm col,5}^{-0.12} \frac{\sqrt{\epsilon_{\rm a}}}{1+\sqrt{\epsilon_{\rm a}}} B_\nu(1.71 \, x^{0.25} T_{\rm col}) \right)^{-m}      
        \right]^{-1/m} & h\nu<3.5 T_{\rm col} \\
        \quad & \quad \\
        1.2\times L_{\nu,\rm BB}(1.11 L_{42.5}^{0.03} T_{5}^{0.18} \times T_{\rm col}) & h\nu>3.5T_{\rm col}
   \end{cases}
\end{multline}
for BSG's. Here, $x=h\nu/T_{\rm col}$, $T_{\rm col} = 5 \, T_{5} \, \rm eV$, and $\epsilon_{\rm a} = [5.5,5.1]\times10^{-3} \, x^{-1.66} T_{\rm col,5}^{-1.098}$. 

\subsection{Validity Time}
Our analytic approximations are valid during 
\begin{equation}
\label{eq:Ats_Appendix}
   \max \left [3 R / c , t_{\rm bo} \right] < t < \min[t_{\rm 0.7 eV}, t_{\rm tr}/a],
\end{equation}
for RSG's, and at
\begin{equation}
   \max \left [2 R / c , t_{\rm bo} \right] < t < \min[t_{\rm 0.8\,  eV}, t_{\rm tr}/a],
\end{equation}
for BSG's. Here,
\begin{equation} \label{eq:tbo_Appendix}
    \begin{split}
        \begin{aligned}
            t_{\rm bo} =& 30 \, R_{13}^{2.16} v_{\rm s*,8.5}^{-1.58} (f_\rho M_0 \kappa_{0.34})^{-0.58} \, \rm sec, \, \, \, \, (RSG) \\
            =& 45 \, R_{13}^{1.90} v_{\rm s*,8.5}^{-1.45} (f_\rho M_0 \kappa_{0.34})^{-0.45} \, \rm sec, \, \, \, \, (BSG)
        \end{aligned}
    \end{split}
\end{equation}
\begin{equation} \label{eq:At_0_7eV_Appendix}
    t_{0.7 \, \rm eV} = 6.86 \, R_{13}^{0.56} {\rm v_{\rm s*,8.5}^{0.16}} \kappa_{0.34}^{-0.61} (f_{\rho}M_0)^{-0.06} \rm  \, days, \, (RSG)
\end{equation}
\begin{equation}
    t_{\rm 0.8\,  eV} = 4.46 R_{13}^{0.52} v_{\rm s*,8.5}^{0.14} (f_{\rho}M_0)^{-0.02}\kappa_{0.34}^{-0.55} \, \rm days,\,(BSG)
\end{equation}
and
\begin{equation}
\label{eq:At_transp}
    \begin{split}
        \begin{aligned}
            t_{\rm tr} &= 19.5 \, \sqrt{\frac{\kappa_{0.34}M_{\rm env,0}} {\rm {v_{\rm s*,8.5}}}}\, \text{days}.
        \end{aligned}
    \end{split}
\end{equation}

The formulae for RSG's were shown to reproduce well the results of numeric calculations over the parameter range: $R=3\times10^{12}-2\times10^{14}$ cm, $E=10^{50}-10^{52}$ erg, total mass $M=2 -40 M_{\odot}$, core and envelope mass ratios $M_{\rm e}/M_{\rm c}=0.3 - 10$, and solar-like metallicity between $Z=0.1-1 \, Z_\odot$. For BSG's, they were shown to reproduce well the results of numeric calculations over the parameter range: $R=10^{12}-3\times10^{13}$ cm, $E=10^{50}-10^{52}$ erg, $M=4 -60 M_{\odot}$, $M_{\rm e}/M_{\rm c}=0.3 - 10$, and solar-like metallicity between $Z=0.1-1 \, Z_\odot$.

\bsp	
\label{lastpage}
\end{document}